\documentclass[preprint,aps,showpacs,amsfonts,epsf]{revtex4}

\input{epsf.tex}

\newcommand{\s}{\sigma}

\newcommand{\be}{\begin{equation}}
\newcommand{\ee}{\end{equation}}
\newcommand{\bea}{\begin{eqnarray}}
\newcommand{\eea}{\end{eqnarray}}
\newcommand{\ba}{\begin{array}}
\newcommand{\ea}{\end{array}}
\def\J#1#2#3#4{{#1} {\bf #2}, #3 (#4)}
\def\PRD{Phys. Rev. D}

\def\JMP{J. Math. Phys.}

\def\CQG{Class. Quantum Grav.}

\begin{document}
\draft
\title{Comment on ``Generalized black diholes''}
\author{V.~S.~Manko$^\dag$ and J. D. Sanabria-G\'omez$^\ddag$}
\address{$^\dag$Departamento de F\'\i sica, Centro de Investigaci\'on y de
Estudios Avanzados del IPN, A.P. 14-740, 07000 M\'exico D.F.,
Mexico\\ $^\ddag$Escuela de F\'isica, Universidad Industrial de
Santander,\\ A.A. 678, Bucaramanga, Colombia}

\begin{abstract}
We show that a recent solution published by Cabrera-Munguia {\it
et al.} is physically inconsistent since the quantity $\s$ it
involves does not have a correct limit $R\to\infty$.
\end{abstract}


\maketitle


In the paper \cite{MRS} we considered a 4-parameter solution from
the Ernst-Manko-Ruiz (EMR) family of equatorially antisymmetric
electrovac spacetimes \cite{EMR} describing a pair of
counter-rotating Kerr-Newman (KN) sources~\cite{NCC} endowed with
opposite electric charges -- a stationary dihole. At the end of
that paper we presented and briefly discussed a generalization of
our 4-parameter model written in physical parametrization to the
case when the two sources, in addition to electric charges, could
also carry arbitrary magnetic opposite charges, the resulting
5-parameter dihole dyonic configuration being defined by the
constant quantity $\s$ of the form (Eq.~(40) of \cite{MRS})
\be \s=\sqrt{M^2-\left(\frac{M^2a^2[(R+2M)^2+4(Q^2+{\mathcal
B}^2)]} {[M(R+2M)+Q^2+{\mathcal B}^2]^2}+Q^2+{\mathcal
B}^2\right)\frac{R-2M}{R+2M}}, \quad a=\frac{J}{M}, \label{sig}
\ee
where $M$, $J$, $a$, $Q$ and ${\mathcal B}$ are, respectively, the
mass, angular momentum, angular momentum per unit mass, electric
charge and magnetic charge of the {\it upper} constituent (the
characteristics of the {\it lower} constituent are correspondingly
$M$, $-J$, $-a$, $-Q$, $-{\mathcal B}$), while $R$ is the
separation distance (see figure~1). Later, after learning about
our results, Cabrera-Munguia {\it et al.} \cite{CLL} have
published a similar representation of the 5-parameter EMR metric
which only slightly differs from ours in the form of $\s$: their
$\s$ is obtainable from (\ref{sig}) via the substitution
\be M^2a^2\equiv J^2 \qquad \mbox{to} \qquad (J-Q{\cal B})^2,
\label{sub} \ee
thus acquiring some additional terms compared to (\ref{sig}).
Therefore, a question naturally arises: which version of the
formula for $\s$ is correct? Unfortunately, the issue of
discrepancy between two $\s$'s was not touched in the paper of
Cabrera-Munguia {\it et al.}, although logically this should have
been the main subject of that paper. Moreover, the authors of
\cite{CLL} made reference to our article exclusively in the
context of the 4-parameter solution, with no mention of our
5-parameter dyonic model. In the present comment we will show that
the expression for $\s$ obtained by Cabrera-Munguia {\it el al.}
with the aid of an ``enhanced'' mass relation is in effect
physically inconsistent.

First of all, we would like to remark that one might naively think
that, since the 5-parameter solutions from \cite{MRS,CLL} differ
in the form of $\s$ only, then the physically incorrect solution
should not satisfy the field equations identically. However, this
is not the case because in the solution construction procedure
employed in the two papers the quantity $\s$ is an arbitrary
constant which may in principle be chosen in the infinite number
of very exotic unphysical ways without violating the field
equations. Hence, some other, less straightforward criteria must
be applied in order to check the physical relevance of the
solutions. Fortunately, the physical inconsistency of formula
(\ref{sig}) after performing the substitution (\ref{sub}) can be
trivially established by considering the limit $R\to\infty$
(infinite separation of the KN sources when the interaction is
absent), which leads to the expression
\be \s=\sqrt{M^2-\frac{(J-Q{\cal B})^2} {M^2}-Q^2-{\mathcal B}^2},
\label{s_lim} \ee
and one can see that the above formula is manifestly different
from the corresponding well-known $\s$ defining the event horizon
of an isolated KN black hole endowed with both electric and
magnetic charges (see, e.g., Eq.~(6.1) of \cite{Car}),
\be \s=\sqrt{M^2-\frac{J^2} {M^2}-Q^2-{\mathcal B}^2}
=\sqrt{M^2-a^2-Q^2-{\mathcal B}^2}. \label{s_is} \ee
To make things worse, the expression (\ref{s_lim}) is not
invariant under the sign change $J\to-J$, $Q\to-Q$, ${\mathcal
B}\to-{\mathcal B}$, the latter transformation converting the term
$(J-Q{\cal B})^2$ into $(J+Q{\cal B})^2$, which clearly violates
the symmetry of the particular two-body problem under
consideration. Moreover, the non-limiting expression for $\s$ must
be also invariant under the above sign change; however, a simple
check shows that the required invariance is absent in the formula
for $\s$ given by Cabrera-Munguia {\it et al.}

Lastly, it might be worth mentioning that in view of the physical
deficiency of the generic 5-parameter solution \cite{CLL} it turns
out that the specific 4-parameter metric earlier presented by
Cabrera-Munguia {\it et al.} \cite{CLL2}, besides its unphysical
character pointed out in \cite{MRS}, must be also inevitably
plagued by an incorrect expression for $\s$, because the latter
was obtained by means of the same ``enhanced'' authors' formula as
the more general $\s$ from \cite{CLL}.

\section*{Acknowledgments}

This work was partially supported by CONACyT of Mexico.


\begin{figure}[htb]
\centerline{\epsfysize=85mm\epsffile{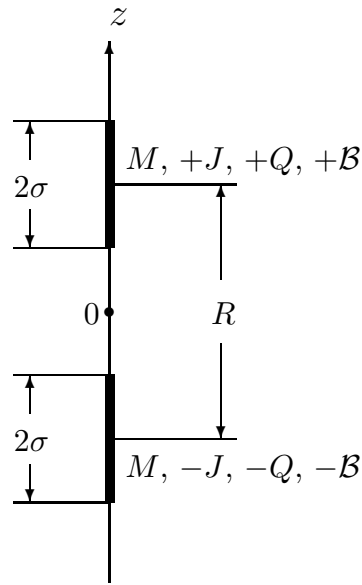}} \caption{Location
of subextreme KN sources on the symmetry axis and the parameters
associated with each source.}
\end{figure}

\end{document}